# Reconfiguring structured light beams using nonlinear metasurfaces

Yun Xu,[1] Jingbo Sun,[1] Jesse Frantz,[2] Mikhail I. Shalaev,[1] Wiktor Walasik,[1] Apra Pandey,[4] Jason D. Myers,[2] Robel Y. Bekele,[3] Alexander Tsukernik,[5] Jasbinder S. Sanghera,[2] Natalia M. Litchinitser[1*]

[1]*Department of Electrical Engineering, the State University of New York, University at Buffalo, Buffalo, NY 14260, USA*

[2]*US Naval Research Laboratory, 4555 Overlook Ave., SW, Washington, DC 20375*

[3]*University Research Foundation, Greenbelt, MD, USA 20770, USA*

[4]*CST of America, LLC, 3979 Freedom Circle, Suite 750, Santa Clara, CA 95054, USA*

[5]*Toronto Nanofabrication Centre, University of Toronto, Toronto, ON M5S 3G4, Canada*

*e-mail: natashal@buffalo.edu

**Abstract:** Ultra-compact, low-loss, fast, and reconfigurable optical components, enabling manipulation of light by light, could open numerous opportunities for controlling light on the nanoscale. Nanostructured all-dielectric metasurfaces have been shown to enable extensive control of amplitude and phase of light in the linear optical regime. Among other functionalities, they offer unique opportunities for shaping the wave front of light to introduce the orbital angular momentum (OAM) to a beam. Such structured light beams bring a new degree of freedom for applications ranging from spectroscopy and micromanipulation to classical and quantum optical communications. To date, reconfigurability or tuning of the optical properties of all-dielectric metasurfaces have been achieved mechanically, thermally, electrically or optically, using phase-change or nonlinear optical materials. However, a majority of demonstrated tuning approaches are either slow or require high optical powers. Arsenic trisulfide ($As_2S_3$) chalcogenide glass offering ultra-fast and large $\chi^{(3)}$ nonlinearity as well as a low two-photon absorption coefficient in the near and mid-wave infrared spectral range, could provide a new platform for the realization of fast and relatively low-intensity reconfigurable metasurfaces. Here, we design and experimentally demonstrate an $As_2S_3$ chalcogenide glass based metasurface that enables reshaping of a conventional Hermite-Gaussian beam with no OAM into an OAM beam at low-intensity levels, while preserves the original beam's amplitude and phase characteristics at high-intensity levels. The proposed metasurface could find applications for a new generation of optical communication systems and optical signal processing.

The discovery of the fact that photons can carry an orbital angular momentum (OAM) opened a new area of optical physics and led to new understanding of a wide range of phenomena [1-4]. The OAM beams possess an azimuthal phase dependence of $\exp(il\phi)$, where the angle $\phi$ is the azimuthal coordinate and the quantized topological charge is denoted by $l \in \mathbb{Z}$ [5]. The OAM beams find important applications, including light-atom interactions [6], manipulation of microscopic objects [7], imaging [8,9], and optical communications [10-13]. Conventionally, OAM beams are generated using spiral phase plates or spatial light modulators [1]. These bulk-optics based devices suit laboratory experiments, but may not be compatible with integrated optics systems that require ultra-compact and flat optical components. Recently, first steps toward the realization of microscale, planar optical components based on liquid crystal technology (q-plates) [14] and optical metasurfaces [16-28] have been made. While many of the first demonstrations of these devices were designed for generating OAM beams with a specific and fixed topological charge, reconfigurability is one of the desired characteristics allowing switching from one charge to another or from the OAM beam to a beam not carrying an OAM. More recently, electrically tunable q-plates [29] as well as mechanically, electrically, thermally and optically tunable metasurfaces have been demonstrated [30-39]. Nevertheless, a majority of the tuning approaches demonstrated to date were not able to simultaneously realize ultra-fast speed and high efficiency.

Nonlinear optical interactions offer a promising way for ultra-fast, picosecond scale, and efficient optical switching, tuning and reconfiguration [40-44]. Recently, several material platforms, including silicon [39,40,43] and GaAs, have been used to realize nonlinear light-matter interactions in optical metasurfaces [44-46]. However, while silicon and GaAs have both high linear refractive indices and Kerr nonlinearities, and low linear losses in the telecommunication range, their two-photon absorption (TPA) coefficients are large, resulting in a low figure of merit. On the contrary, $As_2S_3$ chalcogenide (ChG) glass displays a very good nonlinear figure of merit [47] in both the near-infrared and the mid-wave infrared spectral bands [48,49]. Moreover, its excellent nano-structuring properties, which enable patterning with resolution superior to polymer photoresists, enable a resist-free approach that simplifies fabrication of the $As_2S_3$-based devices to a single-step process [50]. Therefore, we choose the ChG-glass platform to realize a reconfigurable metasurface. In this work, we design, fabricate, and experimentally demonstrate an $As_2S_3$ metasurface capable of reshaping a conventional Hermite-Gaussian beam with no OAM into an OAM beam at low intensity levels, while preserving the original beam's amplitude and phase characteristics at high intensity levels. Such intensity dependent performance is enabled by the Kerr nonlinearity of ChG glass and carefully designed metasurface that relies on guided resonances of the nanoholes made in the ChG thin film.

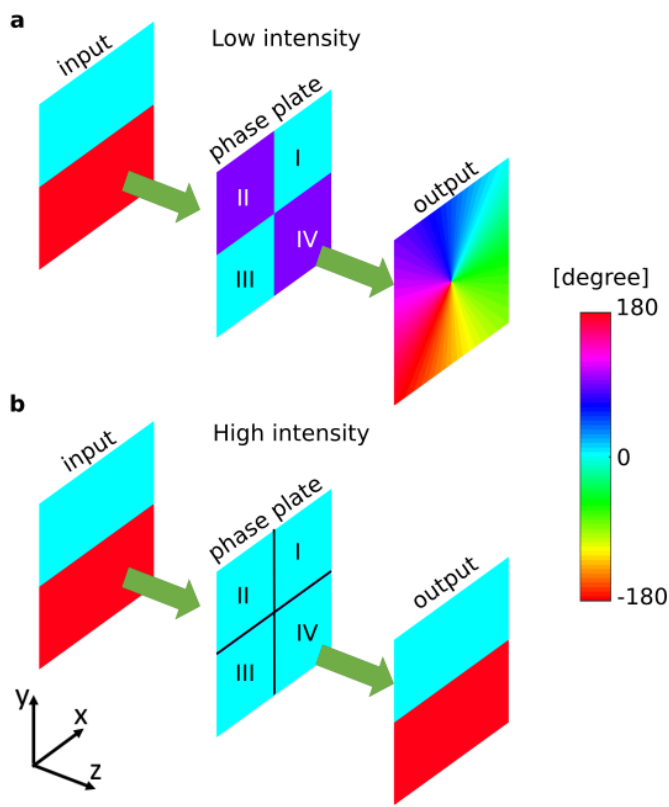

Figure 1. **Working principle of nonlinear metasurface with reconfigurable output beam.** The color maps show the phase of the input and output beams, and the relative phase introduced by the metasurface. **a**, At low intensity, upon transmission through the metasurface, the input Hermite-Gaussian beam acquires a non-uniform phase distribution leading to generation of a beam carrying orbital angular momentum. **b**, At high intensity, the phase difference introduced by the metasurface is uniform; the output beam remains Hermite-Gaussian and does not possess OAM.

The basic principle of operation of the proposed metasurface device is illustrated in Fig. 1. The input Hermite-Gaussian beam is transmitted through the metasurface and acquires different phase distribution depending on the input-light intensity. We design our metasurface such that, in the low-intensity regime, the phase acquired in the even quadrants (II and IV) is larger than for the odd quadrants (I and III). The phase distribution directly after the metasurface has a stepwise profile with the values given in row 2 in Table. 1. After propagation, this step-wise phase change smoothens and becomes a continuously varying phase that characterizes OAM-carrying beams. Therefore, in the low-intensity regime, the HG beam, upon transmission through our metasurface, is transformed into a beam that carries OAM. For a high intensity input beam, the phase introduced by the metasurface is uniform, and the input HG beam maintains its phase and intensity distribution and does not acquire the OAM. The output beam reconfigurability is enabled by the design of the metasurface described in detail below, that uses highly nonlinear ChG glass. Switching between low- and high-intensity regimes allows for dynamic introduction of the OAM in the beam.

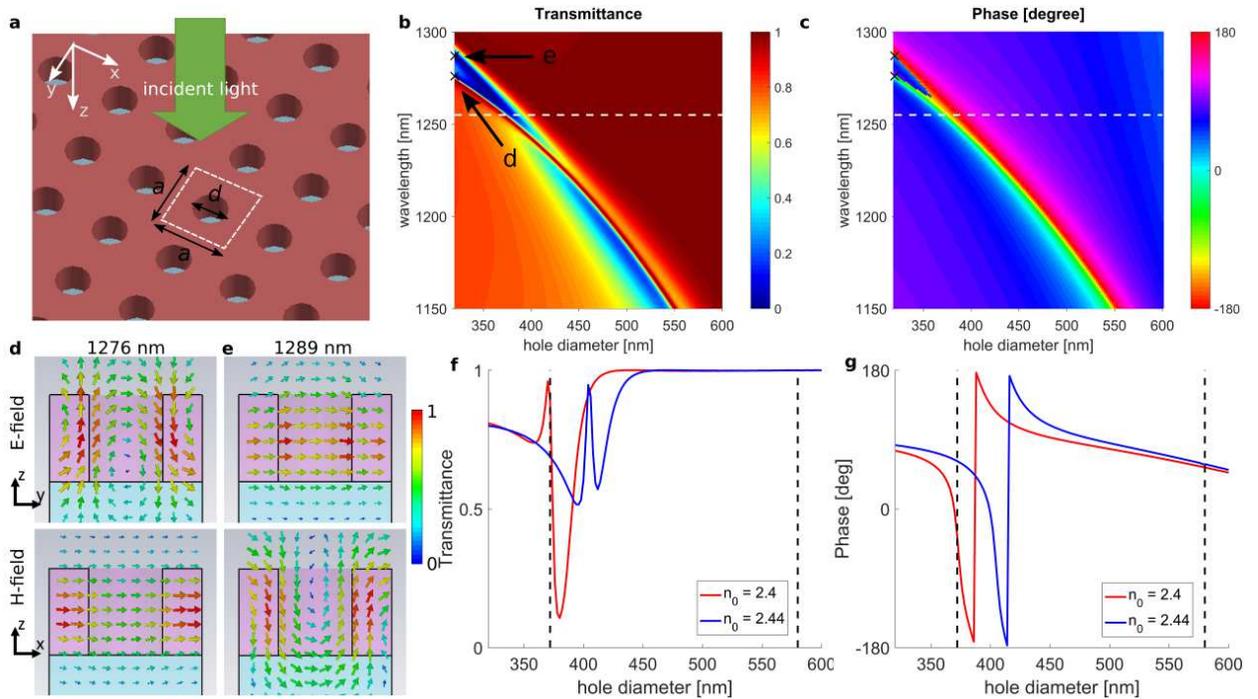

Figure 2. **Design of the nonlinear reconfigurable metasurface. a**, Metasurface consisting of a square lattice of circular holes perforated in a film of arsenic sulfide

($As_2S_3$) ChG glass. The film is deposited on a glass substrate and is covered with a layer of the Poly(methyl methacrylate) (PMMA) which also fills the holes. The geometric parameters of the metasurface are: film thickness, $h = 375$ nm; lattice constant, $a = 660$ nm; and the hole diameter is denoted by $d$. The refractive index of the ChG film is $n_0 = 2.4$. The refractive indices of glass and PMMA are $n_g = n_{PMMA} = 1.5$, making the entire structure symmetric. **b**, **c**, Spectral dependencies of the transmittance and the phase of the transmitted light as a function of the hole diameter, revealing the presence of guided resonances (see Supplementary Materials for more details). The dashed line indicates the working wavelength $\lambda_0 = 1255$ nm. The two black crosses indicate the location of the odd and even resonances whose electric- and magnetic-field distributions are shown in panels **d** and **e**, respectively. **f**, **g** Transmittance and phase of the transmitted light as a function of the hole diameter, for $\lambda = 1255$ nm and two different values of the refractive index of $As_2S_3$. The two dashed lines at $d = 372$ nm and $d = 580$ nm indicate the diameters used in the final design.

The functionality described above is enabled by a ChG-glass-based metasurface illustrated in Fig. 2a. A square array of holes is perforated in a ChG film deposited on a glass substrate. In order to render the structure symmetric, the metasurface is covered with Poly(methyl methacrylate) (PMMA) which also fills the holes. The desired phase difference between the four quadrants of the metasurface is obtained by careful optimization of the geometric parameters. The thickness of the film $h = 375$ nm was a result of the fabrication procedure. The remaining parameters, the lattice constant $a$ and the hole diameter $d$, were selected based on simulations performed using CST MICROWAVE STUDIO. For the structure to operate in the near-infrared, the lattice constant was chosen to be $a = 660$ nm. Figures 2b, 2c show the spectra of the transmittance and the phase of the transmitted light through the metasurface with different hole diameter for the normally incident light. For a hole diameter of $d = 320$ nm, we observe two resonances (odd and even) located at wavelengths $\lambda_o = 1276$ nm and $\lambda_e = 1289$ nm, respectively where the width of the resonance is approximately 5 nm. These resonances originate from the interaction of the incident wave with the modes of the perforated ChG film. For normal incidence, only the doubly degenerate guided resonances can be excited [51], as only for these modes the symmetry of the in-plane electric field components allows for coupling with the incident plane wave [52]. The electric and magnetic field distributions in the metasurface for the light interacting with the odd and even modes are shown in Figs. 2d, 2e.

The relative spectral position of the even and odd resonances can be controlled by changing the hole diameter. For $d = 320$ nm, when the resonances are located apart from each other, we observed two dips in transmittance, at wavelengths $\lambda_o$ and $\lambda_e$. In the vicinity of these wavelengths, the phase of the transmitted light changes rapidly by $180°$. As the diameter of the hole increases, the central wavelengths of the odd and even guided resonances become closer to each other. When the spectral separation between the resonances is smaller than their width, we observe only one dip in transmittance, and the corresponding phase change spans $360°$. For the hole diameter $d = 372$ nm, the resonance is located around the wavelength $\lambda_0 = 1255$ nm. We choose this wavelength to be the operating wavelength for our nonlinear metasurface. Such combination of the hole diameter $d$ and wavelength $\lambda_0$ allows for effective modification of the phase of the transmitted light while small modification of the refractive index $n_{ChG}$ are introduced. The refractive index of the ChG glasses can be efficiently changed as a function of the input light intensity due to their large nonlinear Kerr response described by $n_{ChG} = n_0 + \Delta n = n_0 + n_2 I$. Here, $n_0$ is the linear refractive index of the ChG glass, $n_2 = 7.9 \times 10^{-13}$ cm$^2$/W is the nonlinear coefficient, and $\Delta n$ is the refractive index change of the ChG glass corresponding the input beam intensity $I$.

Figures 2f, 2g show the dependence of the transmittance and the phase of the transmitted light on the hole diameter $d$, at the wavelength $\lambda_0$, for two different values of the linear refractive index of the ChG glass. Here, the modification of the linear refractive index of the ChG glass is assumed to be uniform in the entire film, and is not a result of nonlinear effects. These simulations are used as guidance for our design, and the validity of this approach is confirmed by the full nonlinear results presented in Fig. 3.

As illustrated in Fig. 1, our metasurface design requires the relative phase shift between the odd and even quadrants to be 90° in the low-intensity regime, and 0° in the high-intensity regime. Based on the result shown in Fig. 2g, we observe that these two conditions are fulfilled for the choice of the hole diameter $d_o = 580$ nm and $d_e = 372$ nm, where $d_o$ and $d_e$ denote the hole diameters in odd and even quadrants, respectively. For the structure with $d = d_o$, the resonance is located far away from the operation wavelength, and the structure is insensitive to small change in refractive index. At the same time, for the chosen parameters the transmittance remains above 70%, as seen in Fig. 2f.

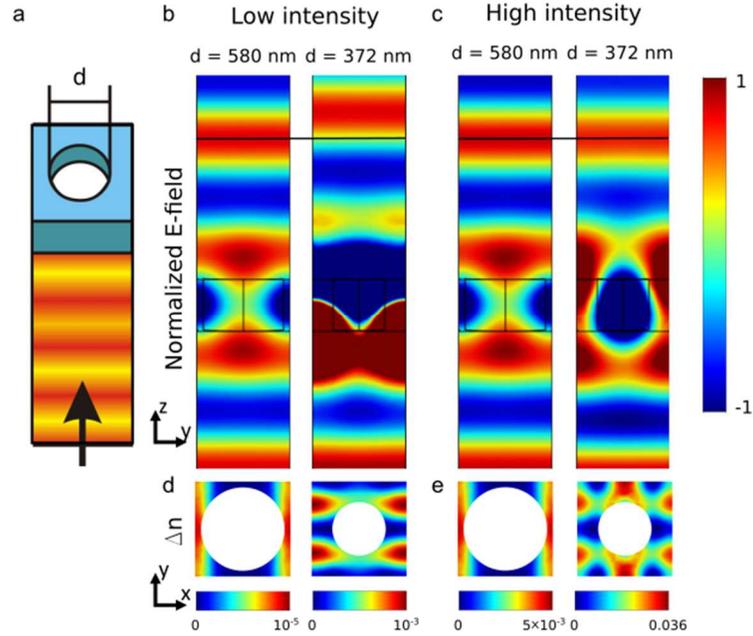

Figure 3. **Nonlinear simulations of a plane-wave propagation through a unit cell with light intensities used in experiment. a**, Schematic of the unit cell used in the simulations showing the propagation direction of the incident $y$-polarized light. **b**, **c**, Electric field distributions in the $x = 0$ plane (center of the unit cell) for low (**b**) and high (**c**) input intensities and for hole diameters $d_o$ (left) and $d_e$ (right) normalized to the maximum input beam intensity. Black horizontal line in the top part of the simulation domain helps visualize the relative phase shift of the transmitted light. The color map for resonant structures is saturated. **d**, **e**, Distribution of the nonlinear index modification in the $z = 0$ plane (center of the ChG film) corresponding to situations illustrated in **b**, **c**.

In order to confirm that the predictions based on uniform linear index changes give us a correct design, we performed nonlinear simulations for the hole arrays with the chosen hole sizes $d_o$ and $d_e$. The results shown in Fig. 3b confirm that for the low intensity (intensity of the incident plane wave equal to 2.4 MW/cm$^2$, which corresponds to the intensity used in the experiment), the phase difference between the odd and even quadrants is equal to 88°. When the intensity is increased to 1.2 GW/ cm$^2$, the waves transmitted through both structures are in phase, as illustrated in Fig. 3c. The maps of the nonlinear refractive index modification $\Delta n$ indicate that in the case of the non-resonant structure ($d = d_o$), the maximum $\Delta n \approx 0.005$, while for the resonant structure ($d = d_e$), the maximum $\Delta n \approx 0.04$. Despite the non-uniform distribution of the refractive index change in the nonlinear case, the phase changes predicted using the results shown in Fig. 2g are in good agreement with the full nonlinear simulations. The nonlinear simulations were performed in COMSOL Multiphysics.

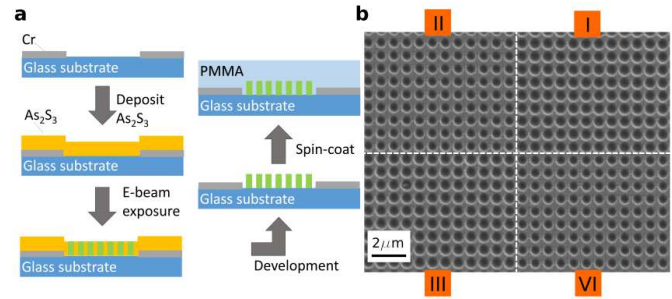

Figure 4. **Fabrication procedure and the fabricated sample. a**, Fabrication process for ChG-based all dielectric metasurfaces involves the following steps: fabrication of a chromium window with size of 150 µm×150 µm; deposition of an As$_2$S$_3$ film; exposure of the As$_2$S$_3$ using electron beam; development of the exposed sample; and spin-coating of a PMMA layer. **b**, Scanning electron microscopy image of the fabricated metasurface with four quadrants, before the layer of PMMA was spin coated.

To realize the functionality described in Fig. 1, a metasurface sample consisting of four quadrants perforated with nano-holes of diameters $d_o$ and $d_e$ was fabricated. The nano-hole arrays designed according to the guided-resonances theory outlines above were patterned in ChG thin film using electron-beam lithography, as shown in Fig. 4. First, chromium windows were prepared on a glass substrate. Then, a ChG film was deposited on top of the chromium windows with thermal deposition. The linear refractive index of the film was measured using spectroscopic ellipsometry to be $n_0 = 2.4$. A solution of diluted MF-319 developer was then used to develop the sample. The developed ChG sample with a thickness of 375 nm is shown in Fig. 4b. Finally, a layer of PMMA was spin coated to fill the holes and cover the fabricated structure in order to obtain a symmetric structure. The fabricated structure contains four quadrants where diagonal quadrants contain holes with the same diameter. The total size of the fabricated metasurface is $132 \times 132\,\mu\text{m}^2$.

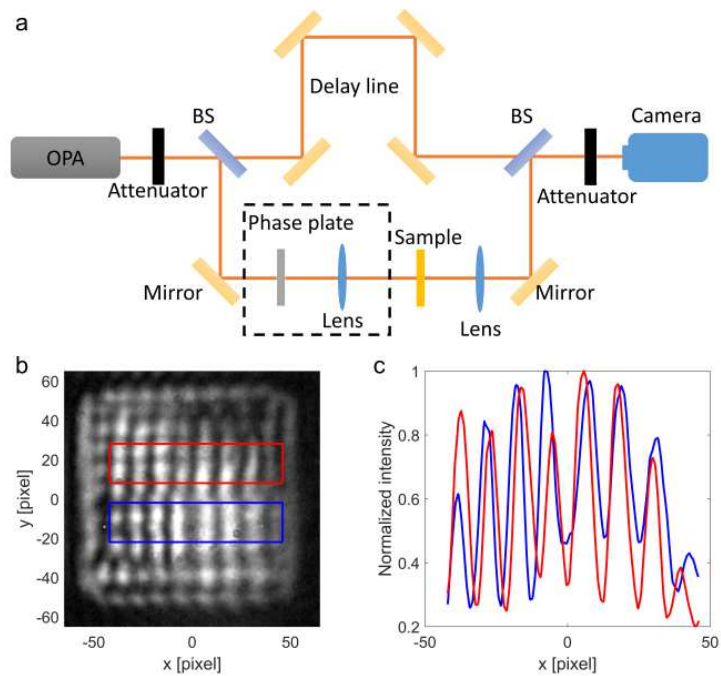

Figure 5. **Metasurface characterization. a**, Schematic of the Mach–Zehnder interferometer with a delay line used to characterize the fabricated metasurface. The phase plate and the lens in the dashed rectangle were added in the second step as described in the text. **b**, Interferogram of the light from the sample plane and the reference beam. **c**, The *y*-averaged intensity distributions in the regions enclosed by rectangles shown in **b** with the colors corresponding to the color of the intensity curve. The relative shift in the peak positions of the red and blue curves allows us to estimate the relative phase shift introduces in the four quadrants, as described in the text.

To characterize the fabricated metasurface, we built a Mach-Zehnder interferometer with a delay line as shown in Fig. 5a. A high-power femtosecond laser system (Coherent Inc.) with optical parametric amplifier generating 100-fs-long pulses with the repetition rate of 1 kHz was used as a light source, with the central wavelength set to 1255 nm. To retrieve the phase difference between the different quadrants of the fabricated metasurface, a low-power collimated Gaussian beam from the laser was split into two: one transmitted through the sample and the other used as a reference beam. The interference pattern of the two beams was then recorded by a camera, and the overlap between the two pulses was controlled by the delay line. The interferogram is shown in Fig. 5b and the phase discontinuities are observed between different quadrants and at the sample outline. The *y*-averaged intensities of the top and bottom parts of the sample are plotted along the *x*-axis of the interferogram in Fig, 5c to determine the phase difference between the neighboring quadrants. For $x<0$, the red curve is shifted toward positive $x$-direction as compared to the blue curve, indicating that the light transmitted through the quadrant II is delayed with respect to the light transmitted by quadrant III. On the contrary, for $x>0$, the red curve is shifted toward negative $x$-direction as compared to the blue curve, indicating that the light transmitted through the quadrant I is advanced with respect to the light transmitted by quadrant IV. The magnitude of this shift is estimated to be a quarter of the fringe period, which corresponds to a 90° phase shift. This confirms the simulation results presented in Fig. 3b.

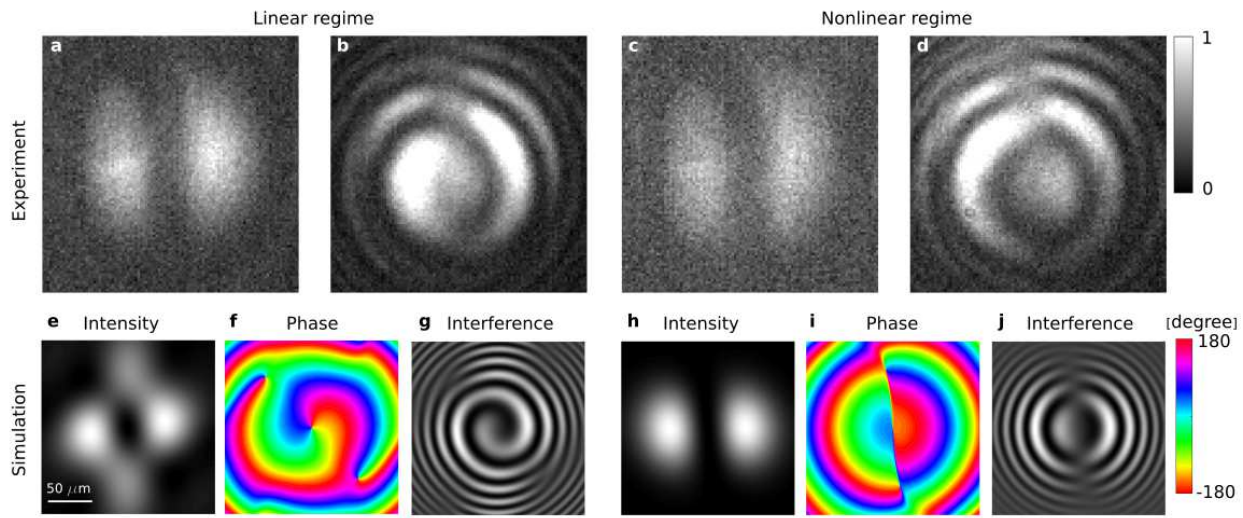

Figure 6. **Experimental and simulation results of a Hermite-Gaussian beam transmitted through the metasurface. a**, Main beam with low intensity after propagating through the metasurface. **c**, Main beam with high intensity after propagating through the metasurface. **b** and **d** are interference of Gaussian beam and beams shown in **a** and **c**, respectively. **e-j,** simulation results of a Hermite-Gaussian beam transmitted through the metasurface with the low refractive index and the high refractive index solved by CST time domain solver. **e-f**, Intensity and phase of the simulated Hermite-Gaussian beam transmitted through the metasurface in linear regime. **g**, Interference of a Gaussian beam and the beam shown in the **e-f**. **h-i**, Intensity and phase of the simulated Hermite-Gaussian beam transmitted through the metasurface in nonlinear regime with higher refractive index. **j**, Interference of a Gaussian beam and the beam shown in the **h-i**.

Once the 90° phase difference between the odd and even quadrants was confirmed in the low-intensity regime, a phase plate was added to the path of the main beam to produce a HG beam. The beam was next focused on the sample using a lens, as shown in the dashed rectangle in Fig. 5a, yielding a beam size of 60 μm. A near-infrared camera was used to capture the main beam transmitted through the metasurface and the interferograms of the main and reference beams. In the low-intensity regime, when the incident beam has the average power of 5 nW, which is not enough to induce a significant nonlinear index change, odd and even quadrants maintain the 90° phase difference as shown in Figs. 2f, 2g. Upon transmission through such metasurface with two quadrants introducing a 90° phase difference, the HG beam acquires an OAM. The transverse intensity distribution of the main beam transmitted through the metasurface is shown in Fig. 6a, and the pattern resulting from the interference of the main beam and the reference Gaussian beam is shown in Fig 6b. The interference pattern reveals has spiral shape, typical for the OAM beams, due to the helical geometry of the phase front. In the nonlinear regime, when the incident beam has the average power of 2 μW (peak intensity around 1.1 GW/cm2), the 90° phase shift in the even quadrants is removed due to the nonlinear response of the metasurface. Therefore, the phase distribution across the entire metasurface becomes uniform and does not affect the phase of the transmitted HG beam. The transverse intensity distribution of the high-intenity main beam transmitted through the metasurface is shown in Fig. 6c, and the corresponding interference pattern with the reference beam is shown in Fig. 6d. The interferogram features semi-circular shapes characteristic for interference of a HG and a Gaussain beams. This result clearly demonstrates the possibility of dynamic introduction of an OAM into a HG beam that can be controlled by the intensity of the input radiation. The transmittance of the metasurface sample in the low and high intensity regimes were measured to be 50% and 53%, respectively.

The simulation results corresponding to the linear and nonlinear regimes are shown in Figs. 6e-g and Figs. 6h-j, respectively. In Fig. 6e-f, we observe a dark beam center and spiral phase characteristic for the beams carrying an OAM. The spiral fringe in Fig. 6g matches the interference experiment shown in Fig. 6b. When the refractive index is increased by 0.04, the beam maintains its HG profile as shown in Fig. 6h-i. The interference pattern in Fig. 6j is corresponding to Fig. 6d. The simulations were performed in CST Microwave Studio with reduced model size of 92.4 μm × 92.4 μm and then propagated in free space for 2.6 mm with the Fast Fourier Transform Beam Propagation Method [53]. The simulations were solved at the wavelength of 1259 nm. Due to the limited computational resources and time, we use a time domain solver and set the mesh to 15 cells per wavelength. With this setting, the 90° phase difference moves 4 nm away from the designed working wavelength.

To conclude, we have experimentally demonstrated that the light transmitted through an all-dielectric metasurface can be reshaped between a beam with no OAM and a beam carrying a well-defined OAM, dependent on the intensity of the input light. The reconfigurability is enabled by the strong Kerr effect in the ChG film, leading to an ultra-fast response. The nonlinear response of the film was enhanced by the use of a resonant design of the metasurface pattern. Highly confined energy density inside the resonant chalcogenide structure results in a reconfigurable metasurface operating at low intensity. Moreover, the metasurface is characterized by a high transmission efficiency around 50% in both the low- and high-intensity modes of operation. Finally, the use of the ChG-glass platform allows for simplification of the fabrication procedure as the $As_2S_3$ glass itself acts as an e-beam resist. This nonlinear reconfigurable metasurface will find applications in ultra-fast and compact optical communication systems, optical signal processing and all optical switching.

**Methods**

**Design.** We used CST Microwave Studio Frequency Solver to design the linear metasurface. The refractive indices of materials were measured using a spectroscopic ellipsometer. The nonlinear simulation is performed by Comsol Multiphysics. The nonlinear coefficient is measured by a home-built Z-scan setup.

**Sample fabrication.** An array of square chromium windows with size $150\mu m \times 150\mu m$ was fabricated on a glass substrate. Then, an $As_2S_3$ film with thickness of 500 nm was deposited on top via thermal evaporation in a Lesker PVD 75 deposition system equipped with a low temperature evaporation source. During deposition, the substrate temperature was maintained at approximately 20° C. Inside each of the Cr windows, the ChG glass was patterned with a square array of holes using electron beam (Vistec EBPG5000+ 100KV) with dosage 10.5 mC/cm$^2$. The area of the pattern is $132\mu m \times 132\mu m$, corresponding to the final sample size. The chromium around the sample plays a double role: (i) it enhances the reflectivity of the substrate enabling automatic sample alignment, and (ii) it increases the sample conductivity allowing to avoid accumulation of charge. After exposure, the sample is immersed in a 1:1 mixture of Microposit MF-319 developer and deionized water for 32 seconds to develop. After development, the exposed area is left and the remaining thickness is 375 nm characterized by atomic force microscopy. Finally, six layers of PMMA was spin-coated on the sample to provide a symmetric refractive index.

**Experiment.** To characterize the fabricated metasurface, we built a Mach-Zehnder interferometer with a delay line as shown in Figure 5a. To generate a Hermite-Gaussian mode, a phase plate was inserted in the main beam path. The phase plate was a glass substrate spin-coated with a layer of S1813 photoresist to delay the beam path for half wavelength. The photoresist on half of the substrate was removed by photolithography. The edge of the photoresist was placed at the center of the main beam and a spatial light filter was placed after the phase plate to pass the Hermite-Gaussian mode.

**Acknowledgments**

This work was supported by the U.S. Office of Naval Research Award N000141613020.

**Author contributions**

Y. X., J. S., and N. M. L. contributed to the initial idea. Y. X., J. S., M. I. S., W. W., and A. P. performed the numerical simulation. J. F., J. D. M, R. Y. B., and J. S. S. prepared the As$_2$S$_3$ ChG film. Y. X., J. S., M. I. S., and A. T. patterned the nanostructures on the samples. Y. X., J. S., M. I. S., W. W., and N. M. L. wrote the manuscript. N. M. L. supervised the work.

**Additional information**

Supplementary information is available.

**Competing financial interests**

The authors declare no competing financial interests.

| Row | Beam | | Phase [°] in quadrant | | | |
|---|---|---|---|---|---|---|
| | | | I | II | III | IV |
| 1 | Input | | 0 | 0 | 180 | 180 |
| 2 | Output | L | 0 | 90 | 180 | 270 |
| 3 | | NL | 0 | 0 | 180 | 180 |

Table 1. Phase of the beam in the four quadrants of the Cartesian coordinate frame before and after transmission of the beam through the metasurface for linear (L) and nonlinear (NL) cases, corresponding to low and high intensity of the input beam, respectively.

**Supplementary information**

In Fig. 2**d**, **e** in the main text, we showed the electric and magnetic field distributions corresponding to two spectrally separated resonances supported by the structure with the hole diameter $d = 320$ nm. To illustrate the origin of the guided resonances and characterize them, in Fig. S1, we show the band structure, transmittance, and phase as functions of frequency for the structure with the same parameters. Figures S1**a**, **b** show the band diagram of the studied structure with the close-up of the spectral range of interest in the vicinity of the center of the first Brillouin zone ($\Gamma$ point). At the $\Gamma$ point, both even and odd the modes of the metasurface can be represented as a superposition of four the modes (with the same parity) of a uniform slab. Normally incident light impinging on the metasurface does not have an in-plane k-vector component and can couple to the modes of the structure located at the $\Gamma$ point. The coupling efficiency depends on the symmetry of the field distribution of the modes and on its overlap integral with the field of the incident wave. As shown in Ref. [1, 2], normally incident light can only couple to the degenerate modes at the $\Gamma$ point. These degenerate odd and even modes are indicated in Fig. S1**b**, and their spectral position matches with the transmittance dips visible in Fig. S1**c**. Figure S1**d** reveals two distinct phase jumps at the resonant frequencies/wavelengths with the amplitudes close to $\pm 180°$.

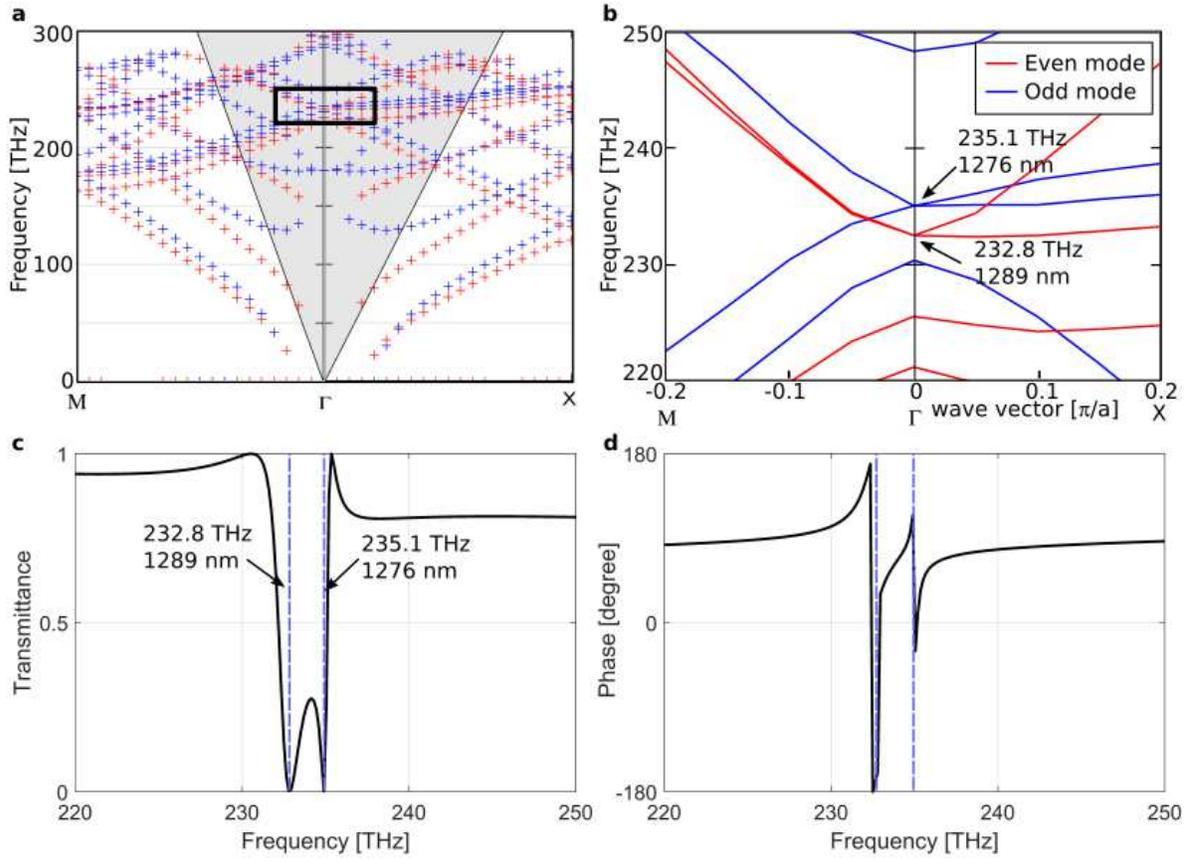

Figure S1. Band structure, transmittance and phase calculated for a square unit cell with periodicity a = 660 nm, thickness $h = 375$ nm, hole diameter $d = 320$ nm and refractive index $n = 2.4$. The refractive index of the surrounding medium (substrate, material inside the hole and on the top of the structure) $n_{env} = 1.5$. **a**. Band structure for the wave vectors along the high symmetry direction MΓ and ΓX in reciprocal space $(k_x, k_y)$. The band structure inside the black rectangle is enlarged in **b**. The degenerate modes are indicated with the arrows and corresponding frequencies. The two degenerate modes at Γ point couple to the electromagnetic wave incident in the direction normal to the surface. Transmittance (**c**) and phase (**d**) for the waves propagating normally to the photonic crystal slab. Two dips in transmittance and two abrupt phase changes in phase are introduced at the wavelengths corresponding to the degenerated modes shown in **b**.

For a single-unit-cell calculations, the frequency-domain results presented in Fig. 2 in the main text and in Figs. S1c-d are obtained using the CST frequency-domain solver with convergence-based mesh refinement [3]. For these simulations, the mesh was refined in the subsequent steps until the difference in calculated transmittance in the two consecutive steps is smaller than 1%. This refinement process allows us to precisely determine the position of the resonances of the structure.

Finally, we found that the size of the simulation domain necessary to model the actual metasurface used in experiments is prohibitively large for the frequency-domain calculations. The simulation results presented in Fig. 6 were obtained using CST time-domain solver. Nevertheless, even in the time-domain simulations, due to the large size of the simulation domain, the limitation of computational resources did not allow us to use an optimum mesh size. The coarser mesh used in the time-domain simulation resulted in small shift in the position of the resonances of the structure. The comparison of the transmittance and phase calculated using the frequency and the time-domain solvers is shown in Fig. S2. It can be seen that, in the time-domain simulation with mesh containing 15 cells per wavelength in the material, the position of the resonance is shifted by 4 nm with respect to the reference frequency-domain result. As a result, the desired 90° phase shift between the quadrants with different hole sizes occurs at a wavelength of 1259 nm. Therefore, the simulations in Figs. 6**e–h** are performed at this wavelength.

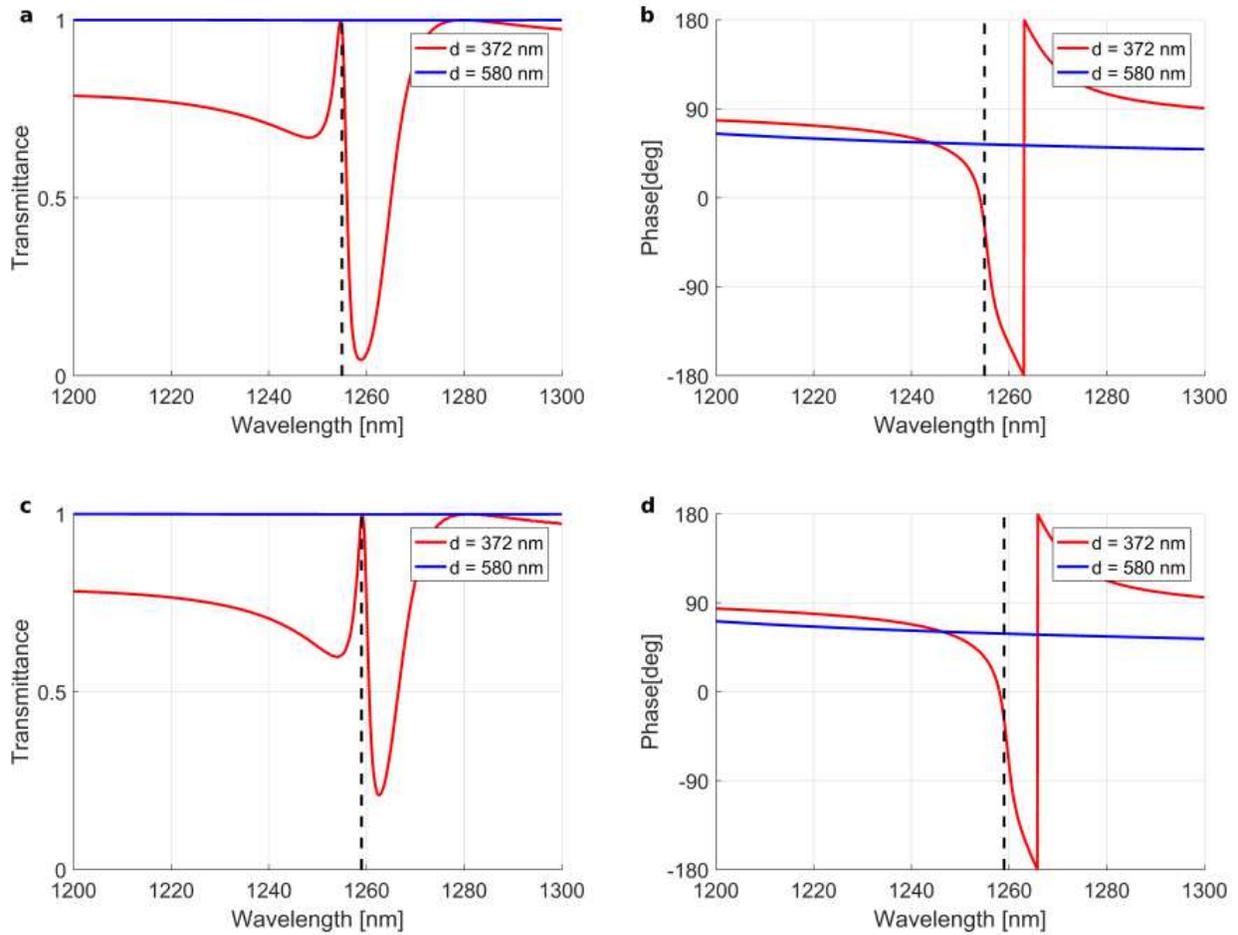

Figure S2. Simulation results of a unit cell with two different diameters *d*. **a**, **b**, Transmittance (**a**) and phase (**b**) obtained using frequency domain solver with mesh refinement described in the text. **c** and **d**, Transmittance (**c**) and phase (**d**) with mesh setting 15 cells per effective wavelength solved by time domain solver.